\RequirePackage{ifpdf}
\ifpdf 
\documentclass[pdftex]{sigma}
\else
\documentclass{sigma}
\fi

\def\Z{{\mathbb Z}} \def\one{\mbox{1 \kern-.59em {\rm l}}}

\def\a{\alpha}

\newcommand{\br}{{\rm BR}}
\newcommand{\amu}{a_\mu}
\newcommand{\amuexp}{a_\mu^{\rm exp}}
\newcommand{\amutheo}{a_\mu^{\rm theo}}

\numberwithin{equation}{section}

\begin{document}


\renewcommand{\PaperNumber}{026}

\FirstPageHeading

\renewcommand{\thefootnote}{$\star$}

\ShortArticleName{Reduction of Couplings: from Finiteness to Fuzzy
Extra Dimensions}

\ArticleName{Unif\/ied Gauge Theories and Reduction of Couplings:\\ from
  Finiteness to Fuzzy Extra Dimensions\footnote{This
paper is a contribution to the Proceedings of the Seventh
International Conference ``Symmetry in Nonlinear Mathematical
Physics'' (June 24--30, 2007, Kyiv, Ukraine). The full collection
is available at
\href{http://www.emis.de/journals/SIGMA/symmetry2007.html}{http://www.emis.de/journals/SIGMA/symmetry2007.html}}}

\Author{Myriam MONDRAG\'ON~$^\dag$ and George ZOUPANOS~$^\ddag$}

\AuthorNameForHeading{M. Mondrag\'on and G. Zoupanos}

\Address{$^\dag$~Inst. de F\'{\i}sica, Universidad Nacional Aut\'onoma de M\'exico, M\'exico}

\EmailD{\href{mailto:myriam@fisica.unam.mx}{myriam@fisica.unam.mx}}

\Address{$^\ddag$~Physics Department, National Technical University,
Athens, Greece}
\EmailD{\href{mailto:george.zoupanos@cern.ch}{george.zoupanos@cern.ch}}

\ArticleDates{Received November 01, 2007, in f\/inal form January 31, 2008; Published online February 23, 2008}

\Abstract{Finite Unif\/ied Theories (FUTs) are $N=1$ supersymmetric Grand Unif\/ied
  Theo\-ries, which can be made all-loop f\/inite, both in the
  dimensionless (gauge and Yukawa couplings) and dimensionful (soft
  supersymmetry breaking terms) sectors.  This remarkable property,
  based on the reduction of couplings at the quantum level, provides a
  drastic reduction in the number of free parameters, which in turn
  leads to an accurate prediction of the top quark mass in the
  dimensionless sector, and predictions for the Higgs boson mass and
  the supersymmetric spectrum in the dimensionful sector.  Here we
  examine the predictions of two such FUTs.  Next we consider gauge
  theories def\/ined in higher dimensions, where the extra dimensions
  form a fuzzy space (a f\/inite matrix manifold).  We reinterpret these
  gauge theories as four-dimensional theories with Kaluza--Klein modes.
  We then perform a~generalized \`a la Forgacs--Manton dimensional
  reduction. We emphasize some striking features emerging such as (i)
  the appearance of non-Abelian gauge theories in four dimensions
  starting from an Abelian gauge theory in higher dimensions, (ii) the
  fact that the spontaneous symmetry breaking of the theory takes
  place entirely in the extra dimensions and (iii) the
  renormaliza\-bility of the theory both in higher as well as in four
  dimensions.  Then reversing the
above approach we present a renormalizable four dimensional $SU(N)$
gauge theory with a suitable multiplet of scalar f\/ields, which via
spontaneous symmetry breaking dynamically  develops extra dimensions
in the form of a fuzzy sphere $S^2_N$. We explicitly f\/ind the tower
of massive Kaluza--Klein modes consistent with an interpretation as
gauge theory on $M^4 \times S^2$, the scalars being interpreted as
gauge f\/ields on $S^2$. Depending on the parameters of the model the
low-energy gauge group can be $SU(n)$, or broken further to $SU(n_1)
\times SU(n_2) \times U(1)$. Therefore the second picture justif\/ies
the f\/irst one in a renormalizable framework but in addition has the
potential to reveal new aspects of the theory.}


\Classification{81T60; 81V22; 83E15; 81T75; 54A40}

\renewcommand{\thefootnote}{\arabic{footnote}}
\setcounter{footnote}{0}

\section {Introduction}

The theoretical ef\/forts to establish a deeper understanding of
Nature have led to very interes\-ting frameworks such as String
theories and Non-commutative Geometry both of which aim to
describe physics at the Planck scale. Looking for the origin of
the idea that coordinates might not commute we might have to go
back to the days of Heisenberg. In the recent years the birth of
such speculations can be found in~\cite{Connes,Madore}. In
the spirit of Non-commutative Geometry also particle models with
non-commutative gauge theory were explored~\cite{Connes:1990qp}
(see also~\cite{Martin:1996wh}), \cite{DV.M.K., M.}. On the other
hand the present intensive research has been triggered by the
natural realization of non-commutativity of space in the string
theory context of D-branes in the presence of a~constant
background antisymmetric f\/ield \cite{Connes:1997cr}. After the
work of Seiberg and Witten~\cite{SW}, where a~map (SW~map) between
non-commutative and commutative gauge theories has been described,
there has been a lot of activity also in the construction of
non-commutative phenomenological Lagrangians, for example various
non-commutative standard model like Lagrangians have been proposed
\cite{Chaic, SM}\footnote{These SM actions are mainly considered
as ef\/fective actions because they are not renormalizable. The
ef\/fective action interpretation is consistent with the SM in~\cite{SM} being anomaly free \cite{martin}. Non-commutative
phenomenology has been discussed in~\cite{phenomenology}.}. In
particular in~\cite{SM}, following the SW map methods
developed in~\cite{Jurco:2000ja}, a~non-commutative standard
model with $SU(3)\times SU(2)\times U(1)$ gauge group has been
presented. These non-commutative models represent interesting
generalizations of the SM and hint at possible new physics.
However they do not address the usual problem of the SM, the
presence of a~plethora of free parameters mostly related to the ad
hoc introduction of the Higgs and Yukawa sectors in the theory. At
this stage it is worth recalling that various schemes, with the
Coset Space Dimensional Reduction (CSDR)
\cite{Forgacs:1979zs,Kapetanakis:hf,Kubyshin:vd,Bais:td} being
pioneer, were suggesting that a unif\/ication of the gauge and Higgs
sectors can be achieved in higher dimensions. Moreover the
addition of fermions in the higher-dimensional gauge theory leads
naturally after CSDR to Yukawa couplings in four dimensions. In
the successes of the CSDR scheme certainly should be added the
possibility to obtain chiral theories in four dimensions
\cite{Farakos:1986sm,Chapline:wy,Barnes:ea,Hanlon} as well as
softly broken supersymmetric or non-supersymmetric theories
starting from a supersymmetric gauge theory def\/ined in higher
dimensions \cite{Manousselis:2001re}.

The original plan of this paper was to present an overview covering
the following subjects:
\begin{enumerate}\itemsep=0pt
\item[a)] Quantum Reduction of Couplings and Finite Unif\/ied Theories;
\item[b)] Classical Reduction of Couplings and Coset Space Dimensional Reduction;
\item[c)] Renormalizable Unif\/ied Theories from Fuzzy Higher Dimensions \cite{fuzzy1}.
\end{enumerate}

The aim was to present an unif\/ied description of our current attempts
to reduce the free parameters of the Standard Model by using Finite
Unif\/ication and extra dimensions.  However we will cover only the
f\/irst and the third subjects given the fact that there exists extensive
reviews covering a major part of the second one
\cite{Kapetanakis:hf,Kubyshin:vd}.  These two topics represent dif\/ferent
attempts at reduction of couplings, on one hand the Finite Unif\/ied Theories
showing promising models with good phenomenology, on the other hand,
the Unif\/ied Theories from Fuzzy Higher Dimensions combining dimensional
reduction and reduction of couplings in a  renormalizable theory.

\section[Reduction of Couplings and Finiteness in $N=1$ SUSY Gauge
  Theories]{Reduction of Couplings and Finiteness\\ in $\boldsymbol{N=1}$ SUSY Gauge
  Theories}

Finite Unif\/ied Theories are $N=1$ supersymmetric Grand Unif\/ied
Theories (GUTs) which can be made f\/inite even to all-loop orders,
including the soft supersymmetry breaking sector.  The method to
construct GUTs with reduced independent parameters
\cite{zoup-kmz1,zoup-zim1} consists of searching for renormalization
group invariant (RGI) relations holding below the Planck scale, which
in turn are preserved down to the GUT scale. Of particular interest is
the possibility to f\/ind RGI relations among couplings that guarantee
f\/initenes to all-orders in perturbation theory
\cite{zoup-lucchesi1,zoup-ermushev1}.  In order to achieve the latter
it is enough to study the uniqueness of the solutions to the one-loop
f\/initeness conditions \cite{zoup-lucchesi1,zoup-ermushev1}.  The
constructed {\it finite unified} $N=1$ supersymmetric $SU(5)$ GUTs, using
the above tools, predicted correctly from the dimensionless sector
(gauge-Yukawa unif\/ication), among others, the top quark mass
\cite{Kapetanakis:vx}.  The search for RGI relations and f\/initeness has
been extended to the soft supersymmetry breaking sector (SSB) of these
theories \cite{zoup-kmz2,zoup-jack2}, which involves parameters of
dimension one and two.  Eventually, the full theories can be made
all-loop f\/inite and their predictive power is extended to the Higgs
sector and the supersymmetric spectrum (s-spectrum).

Here let us review the main points and ideas concerning the {\it reduction
of couplings} and {\it finiteness} in $N=1$ supersymmetric theories.
A RGI relation among couplings $g_i$, $ \Phi (g_1,\dots,g_N) =0, $ has to
satisfy the partial dif\/ferential equation $ \mu\, d \Phi /d \mu=
\sum\limits_{i=1}^{N}  \beta_{i} \partial \Phi /\partial g_{i} = 0$,  where $\beta_i$ is the
$\beta$-function of~$g_i$.  There exist ($N-1$) independent $\Phi$'s, and
f\/inding the complete set of these solutions is equivalent to solve the
so-called reduction equations (REs) \cite{zoup-zim1}, $ \beta_{g}  (d
g_{i}/d g) =\beta_{i}$, $i=1,\dots,N, $ where $g$ and $\beta_{g}$ are the
primary coupling and its $\beta$-function.  Using all the $(N-1) \Phi$'s
to impose RGI relations, one can in principle express all the
couplings in terms of a single coup\-ling~$g$.  The complete reduction,
which formally preserves perturbative renormalizability, can be
achieved by demanding a power series solution, whose uniqueness can be
investigated at the one-loop level.

Finiteness can be understood by considering a chiral, anomaly free,
$N=1$ globally supersymmetric gauge theory based on a group G with
gauge coupling constant $g$. The superpotential of the theory is given
by
\begin{gather}
 W= \tfrac{1}{2} m^{ij} \Phi_{i} \Phi_{j}+
\tfrac{1}{6} C^{ijk}  \Phi_{i} \Phi_{j} \Phi_{k},
\label{supot}
\end{gather}
where $m^{ij}$ (the mass terms) and $C^{ijk}$ (the Yukawa couplings)
are gauge invariant tensors and the matter f\/ield $\Phi_{i}$ transforms
according to the irreducible representation $R_{i}$ of the gauge group~$G$.

 The one-loop $\beta$-function of the gauge
coupling $g$ is given by
\begin{gather*}
\beta^{(1)}_{g} = \frac{d g}{d t} =
\frac{g^3}{16\pi^2}\left[ \sum_{i} T(R_{i})-3 C_{2}(G)\right],
\end{gather*}
where $T(R_{i})$ is the Dynkin index of $R_{i}$ and $C_{2}(G)$
 is the
quadratic Casimir of the adjoint representation of the
gauge group $G$. The $\beta$-functions of
$C^{ijk}$,
by virtue of the non-renormalization theorem, are related to the
anomalous dimension matrix $\gamma^j_i$ of the matter f\/ields
$\Phi_{i}$ as:
\begin{gather*}
\beta_{C}^{ijk}=\frac{d}{dt}\,C^{ijk}
=C^{ijp}
\sum_{n=1}\frac{1}{(16\pi^2)^n}\,\gamma_{p}^{k(n)} +(k
\leftrightarrow i) +(k\leftrightarrow j).
\end{gather*}
At one-loop level $\gamma^j_i$ is given by
\begin{gather*}
\gamma_i^{j(1)}=\tfrac{1}{2}C_{ipq} C^{jpq}-2 g^2 C_{2}(R_{i})\delta_i^j,
\end{gather*}
where $C_{2}(R_{i})$ is the quadratic Casimir of the representation
$R_{i}$, and $C^{ijk}=C_{ijk}^{*}$.

All the one-loop $\beta$-functions of the theory vanish if the
$\beta$-function of the gauge coupling $\beta_g^{(1)}$, and the
anomalous dimensions $\gamma_i^{j(1)}$,
vanish, i.e.
\begin{gather}
\sum _i T (R_i) = 3 C_2(G) ,\qquad
\tfrac{1}{2}C_{ipq} C^{jpq} = 2\delta _i^j g^2  C_2(R_i) .
\label{zoup-fini}
\end{gather}

The conditions for f\/initeness for $N=1$ f\/ield theories with $SU(N)$ gauge
symmetry are discussed in \cite{Rajpoot:1984zq}, and the
analysis of the anomaly-free and no-charge renormalization
requirements for these theories can be found in \cite{Rajpoot:1985aq}.
A very interesting result is that the conditions (\ref{zoup-fini}) are
necessary and suf\/f\/icient for f\/initeness at
the two-loop level \cite{Parkes:1984dh}.

The one- and two-loop f\/initeness conditions (\ref{zoup-fini}) restrict
considerably the possible choices of the irreducible
representations $R_i$ for a given group $G$ as well as the Yukawa
couplings in the superpotential (\ref{supot}).  Note in particular
that the f\/initeness conditions cannot be applied to the supersymmetric
standard model (SSM), since the presence of a $U(1)$ gauge group is
incompatible with the condition (\ref{zoup-fini}), due to
$C_2[U(1)]=0$.  This leads to the expectation that f\/initeness should
be attained at the grand unif\/ied level only, the SSM being just the
corresponding low-energy, ef\/fective theory.

The f\/initeness conditions impose relations between gauge and Yukawa
couplings.  Therefore, we have to guarantee that such relations
leading to a reduction of the couplings hold at any renormalization
point.  The necessary, but also suf\/f\/icient, condition for this to
happen is to require that such relations are solutions to the
reduction equations (REs) to all orders.  The all-loop order
f\/initeness theorem of~\cite{zoup-lucchesi1} is based on: (a) the
structure of the supercurrent in $N=1$ SYM and on (b) the
non-renormalization properties of $N=1$ chiral
anomalies~\cite{zoup-lucchesi1}.  Alternatively, similar results can be obtained~\cite{zoup-ermushev1,zoup-strassler} using an analysis of the all-loop
NSVZ gauge beta-function~\cite{zoup-novikov1}.

\section[Soft supersymmetry breaking and finiteness]{Soft supersymmetry breaking and f\/initeness}

The above described method of reducing the dimensionless couplings has
been extended \cite{zoup-jack2,zoup-kmz2} to the soft supersymmetry
breaking (SSB) dimensionful parameters of $N=1$ supersymmetric
theories.  Recently very interesting progress has been made
\cite{zoup-avdeev1,zoup-kkz,zoup-yamada1,zoup-hisano1,zoup-kkk1,zoup-delbourgo1,zoup-girardello1,zoup-jack4,zoup-kkmz1}
concerning the renormalization properties of the SSB parameters, based
conceptually and technically on the work of~\cite{zoup-yamada1}.
In this work the powerful supergraph method \cite{zoup-delbourgo1} for
studying supersymmetric theories has been applied to the softly broken
ones by using the ``spurion'' external space-time independent
superf\/ields \cite{zoup-girardello1}.  In the latter method a softly
broken supersymmetric gauge theory is considered as a supersymmetric
one in which the various parameters such as couplings and masses have
been promoted to external superf\/ields that acquire ``vacuum
expectation values''. Based on this method the relations among the
soft term renormalization and that of an unbroken supersymmetric
theory have been derived. In particular the $\beta$-functions of the
parameters of the softly broken theory are expressed in terms of
partial dif\/fe\-ren\-tial ope\-ra\-tors involving the dimensionless parameters
of the unbroken theory. The key point in the strategy of~\cite{zoup-avdeev1,zoup-kkz,zoup-yamada1,zoup-hisano1,zoup-kkk1,zoup-delbourgo1,zoup-girardello1,zoup-jack4,zoup-kkmz1}
in solving the set of
coupled dif\/ferential equations so as to be able to express all
parameters in a RGI way, was to transform the partial dif\/fe\-ren\-tial
operators involved to total derivative operators~\cite{zoup-avdeev1}.
It is indeed possible to do this on the RGI surface which is def\/ined
by the solution of the reduction equations.  In addition it was found
that RGI SSB scalar masses in gauge-Yukawa unif\/ied models satisfy a
universal sum rule at one-loop~\cite{zoup-kkk1}. This result was
generalized to two-loops for f\/inite theories \cite{zoup-kkmz1}, and
then to all-loops for general gauge-Yukawa and Finite Unif\/ied Theories
\cite{zoup-kkz}.

In order to obtain a feeling of some of the above
results, consider the superpotential given by~\eqref{supot} along with the
Lagrangian for SSB terms
\begin{gather*}
-{\cal L}_{\rm SB} = \tfrac{1}{6}
 h^{ijk} \phi_i \phi_j \phi_k +
\tfrac{1}{2}  b^{ij} \phi_i \phi_j
+ \tfrac{1}{2}  (m^2)^{j}_{i} \phi^{*\,i} \phi_j+ \tfrac{1}{2} M\lambda
\lambda+\mbox{H.c.},
\end{gather*}
where the $\phi_i$ are the scalar parts of the
chiral superf\/ields $\Phi_i$, $\lambda$ are the gauginos and $M$ their
unif\/ied mass.  Since only f\/inite theories are considered here,
it is assumed that the gauge group is a simple group and the
one-loop $\beta$-function of the gauge coupling $g$ vanishes. It is also
assumed that the reduction equations admit power series solutions of
the form
\begin{gather*}
C^{ijk} = g \sum_{n=0} \rho^{ijk}_{(n)} g^{2n}.
\end{gather*}
According
to the f\/initeness theorem \cite{zoup-lucchesi1}, the theory is then f\/inite
to all-orders in perturbation theory, if, among others, the one-loop
anomalous dimensions $\gamma_{i}^{j(1)}$ vanish.  The one- and two-loop
f\/initeness for $h^{ijk}$ can be achieved by \cite{zoup-jack1}
\begin{gather*}
 h^{ijk} = -M C^{ijk}+\cdots =-M
\rho^{ijk}_{(0)} g+O(g^5) .
\end{gather*}

An additional constraint in the SSB sector up to two-loops
\cite{zoup-kkmz1}, concerns the soft scalar masses as follows
\begin{gather}
\frac{( m_{i}^{2}+m_{j}^{2}+m_{k}^{2} )}{M M^{\dag}} =
1+\frac{g^2}{16 \pi^2} \Delta^{(2)}
+O(g^4)
\label{zoup-sumr}
\end{gather}
for $i$, $j$, $k$ with $\rho^{ijk}_{(0)} \neq 0$, where $\Delta^{(2)}$ is
the two-loop correction
\begin{gather*}
\Delta^{(2)} =  -2\sum_{l} [(m^{2}_{l}/M M^{\dag})-(1/3)] T(R_l),
\end{gather*}
which vanishes for the
universal choice \cite{zoup-jack1}, i.e.\ when all the soft scalar
masses are the same at the unif\/ication point.

If we know higher-loop $\beta$-functions explicitly, we can follow the same
procedure and f\/ind higher-loop RGI relations among SSB terms.
However, the $\beta$-functions of the soft scalar masses are explicitly
known only up to two loops.
In order to obtain higher-loop results, we need something else instead of
knowledge of explicit $\beta$-functions, e.g.\ some relations among
$\beta$-functions.

The recent progress made using the spurion technique
\cite{zoup-delbourgo1,zoup-girardello1} leads to
the following  all-loop relations among SSB $\beta$-functions
\cite{zoup-avdeev1,zoup-kkz,zoup-yamada1,zoup-hisano1,zoup-kkk1,zoup-delbourgo1,zoup-girardello1,zoup-jack4,zoup-kkmz1}
\begin{gather*}
\beta_M  =  2{\cal O}\left(\frac{\beta_g}{g}\right) ,\\
\beta_h^{ijk} = \gamma^i{}_lh^{ljk}+\gamma^j{}_lh^{ilk}
+\gamma^k{}_lh^{ijl} -2\gamma_1^i{}_lC^{ljk}
-2\gamma_1^j{}_lC^{ilk}-2\gamma_1^k{}_lC^{ijl} ,\\
(\beta_{m^2})^i{}_j  = \left[ \Delta
+ X \frac{\partial}{\partial g}\right]\gamma^i{}_j,\\
{\cal O}  = \left(Mg^2{\frac{\partial}{\partial g^2}}
-h^{lmn}\frac{\partial}{\partial C^{lmn}}\right),\\
\Delta  =  2{\cal O}{\cal O}^* +2|M|^2 g^2\frac{\partial}{\partial g^2}  +\tilde{C}_{lmn}
\frac{\partial}{\partial C_{lmn}} +\tilde{C}^{lmn}\frac{\partial}{\partial C^{lmn}} ,
\end{gather*}
where $(\gamma_1)^i{}_j={\cal O}\gamma^i{}_j$,
$C_{lmn} = (C^{lmn})^*$, and
\begin{gather*}
\tilde{C}^{ijk}=
(m^2)^i{}_lC^{ljk}+(m^2)^j{}_lC^{ilk}+(m^2)^k{}_lC^{ijl} .
\end{gather*}
It was also found \cite{zoup-jack4}  that the relation
\begin{gather*}
h^{ijk} = -M (C^{ijk})'
\equiv -M \frac{d C^{ijk}(g)}{d \ln g} ,
\end{gather*}
among couplings is all-loop RGI. Furthermore, using the all-loop gauge
$\beta$-function of Novikov  et al.~\cite{zoup-novikov1} given
by
\begin{gather*}
\beta_g^{\rm NSVZ} =
\frac{g^3}{16\pi^2}
\left[ \frac{\sum_l T(R_l)(1-\gamma_l /2)
-3 C(G)}{ 1-g^2C(G)/8\pi^2}\right],
\end{gather*}
it was found the all-loop RGI sum rule \cite{zoup-kkz},
\begin{gather*}
m^2_i+m^2_j+m^2_k =
|M|^2 \left\{
\frac{1}{1-g^2 C(G)/(8\pi^2)}\frac{d \ln C^{ijk}}{d \ln g}
+\frac{1}{2}\frac{d^2 \ln C^{ijk}}{d (\ln g)^2}\right\}\nonumber\\
\phantom{m^2_i+m^2_j+m^2_k =}{}
 +\sum_l
\frac{m^2_l T(R_l)}{C(G)-8\pi^2/g^2}
\frac{d \ln C^{ijk}}{d \ln g}.
\end{gather*}
In addition
the exact-$\beta$-function for $m^2$
in the NSVZ scheme has been obtained \cite{zoup-kkz} for the f\/irst time and
is given by
\begin{gather*}
\beta_{m^2_i}^{\rm NSVZ} =\left[
|M|^2 \left\{
\frac{1}{1-g^2 C(G)/(8\pi^2)}\frac{d }{d \ln g}
+\frac{1}{2}\frac{d^2 }{d (\ln g)^2}\right\}\right. \nonumber\\
 \left. \phantom{\beta_{m^2_i}^{\rm NSVZ} =}{} +\sum_l
\frac{m^2_l T(R_l)}{C(G)-8\pi^2/g^2}
\frac{d }{d \ln g} \right] \gamma_{i}^{\rm NSVZ}.
\end{gather*}

\section[Finite Unified Theories]{Finite Unif\/ied Theories}

 The f\/irst
one- and two-loop $SU(5)$ f\/inite model was presented in~\cite{Jones:1984qd}.  Here we will examine Finite Unif\/ied Theories
with $SU(5)$ gauge group, where the reduction of couplings has been
applied to the third generation of quarks and leptons.  An extension
to three families, and the generation of quark mixing angles and
masses in Finite Unif\/ied Theories has been addressed in~\cite{Babu:2002in}, where several examples are given. These extensions
are not considered here.  Realistic Finite Unif\/ied Theories based on
product gauge groups, where the f\/initeness implies three generations
of matter, have also been studied~\cite{Ma:2004mi}.

 A predictive gauge-Yukawa
unif\/ied $SU(5)$ model which is f\/inite to all orders, in addition to
the requirements mentioned already, should also have the following
properties:
\begin{enumerate}\itemsep=0pt
\item
One-loop anomalous dimensions are diagonal,
i.e.,  $\gamma_{i}^{(1)\,j} \propto \delta^{j}_{i} $.
\item Three fermion generations, in the irreducible representations
  $\overline{\bf 5}_{i}$, ${\bf 10}_i$ $(i=1,2,3)$, which obviously should
  not couple to the adjoint ${\bf 24}$.
\item The two Higgs doublets of the MSSM should mostly be made out of a
pair of Higgs quintet and anti-quintet, which couple to the third
generation.
\end{enumerate}

In the following we discuss two versions of the all-order f\/inite
model.    The model of~\cite{Kapetanakis:vx}, which
will be labeled ${\bf A}$, and a slight variation of this model
(labeled ${\bf B}$), which can also be obtained from the class of the
models suggested in~\cite{zoup-avdeev1} with a modif\/ication to
suppress non-diagonal anomalous dimensions.

The  superpotential which describes the two models
takes the form \cite{Kapetanakis:vx,zoup-kkmz1}
\begin{gather*}
W = \sum_{i=1}^{3}\left[\tfrac{1}{2}g_{i}^{u}
 {\bf 10}_i{\bf 10}_i H_{i}+
g_{i}^{d} {\bf 10}_i \overline{\bf 5}_{i}
\overline{H}_{i}\right]
+g_{23}^{u} {\bf 10}_2{\bf 10}_3 H_{4}
  +g_{23}^{d} {\bf 10}_2 \overline{\bf 5}_{3}
\overline{H}_{4}\nonumber\\
\phantom{W =}{} +
g_{32}^{d} {\bf 10}_3 \overline{\bf 5}_{2}
\overline{H}_{4}
+\sum_{a=1}^{4}g_{a}^{f} H_{a}
{\bf 24} \overline{H}_{a}+
\frac{g^{\lambda}}{3} ({\bf 24})^3,
\end{gather*}
where
$H_{a}$ and $\overline{H}_{a}$ $(a=1,\dots,4)$
stand for the Higgs quintets and anti-quintets.

The non-degenerate and isolated solutions to $\gamma^{(1)}_{i}=0$ for
the models $\{ {\bf A},{\bf B} \}$ are:
\begin{gather} (g_{1}^{u})^2
=\left\{\tfrac{8}{5},\tfrac{8}{5} \right\}g^2, \qquad (g_{1}^{d})^2
=\left\{\tfrac{6}{5},\tfrac{6}{5}\right\}g^2,\qquad
(g_{2}^{u})^2=(g_{3}^{u})^2=\left\{\tfrac{8}{5},\tfrac{4}{5}\right\}g^2,\nonumber\\
(g_{2}^{d})^2 =(g_{3}^{d})^2=\left\{\tfrac{6}{5},\tfrac{3}{5}\right\}g^2,\qquad
(g_{23}^{u})^2 =\left\{0,\tfrac{4}{5}\right\}g^2,\qquad
(g_{23}^{d})^2=(g_{32}^{d})^2=\left\{0,\tfrac{3}{5}\right\}g^2,
\nonumber\\
(g^{\lambda})^2 =\tfrac{15}{7}g^2,\qquad (g_{2}^{f})^2
=(g_{3}^{f})^2=\left\{0,\tfrac{1}{2}\right\}g^2,\qquad
(g_{1}^{f})^2 = 0,\qquad
(g_{4}^{f})^2=\{1,0\}g^2.\label{zoup-SOL5}
\end{gather}
According to the theorem of~\cite{zoup-lucchesi1} these models are f\/inite to all orders.  After the
reduction of couplings the symmetry of $W$ is enhanced~\cite{Kapetanakis:vx,zoup-kkmz1}.

The main dif\/ference of the models ${\bf A}$ and ${\bf B}$ is that two
pairs of Higgs quintets and anti-quintets couple to the ${\bf 24}$ for
${\bf B}$ so that it is not necessary to mix them with $H_{4}$ and
$\overline{H}_{4}$ in order to achieve the triplet-doublet splitting
after the symmetry breaking of $SU(5)$.  Therefore, the solutions
of equation~(\ref{zoup-fini}) for the Yukawa couplings are dif\/ferent, as can
be seen from equation~(\ref{zoup-SOL5}), which ref\/lects in the
phenomenology, as we will see in the next section.

In the dimensionful sector, the sum rule gives us the following
boundary conditions at the GUT scale \cite{zoup-kkmz1}:
\begin{gather*}
m^{2}_{H_u}+
2  m^{2}_{{\bf 10}} =
m^{2}_{H_d}+ m^{2}_{\overline{{\bf 5}}}+
m^{2}_{{\bf 10}}=M^2\qquad \mbox{for}\quad {\bf A} ;\\
m^{2}_{H_u}+
2  m^{2}_{{\bf 10}} =M^2,\qquad
m^{2}_{H_d}-2m^{2}_{{\bf 10}}=-\frac{M^2}{3},\qquad
m^{2}_{\overline{{\bf 5}}}+
3m^{2}_{{\bf 10}} = \frac{4M^2}{3}\qquad \mbox{for}\quad {\bf B},
\end{gather*}
where we use as  free parameters
$m_{\overline{{\bf 5}}}\equiv m_{\overline{{\bf 5}}_3}$ and
$m_{{\bf 10}}\equiv m_{{\bf 10}_3}$
for the model ${\bf A}$, and
$m_{{\bf 10}}\equiv m_{{\bf 10}_3}$  for ${\bf B}$, in addition to $M$.

\section{Predictions of low energy parameters}

Since the gauge symmetry is spontaneously broken below $M_{\rm GUT}$,
the f\/initeness conditions do not restrict the renormalization properties
at low energies, and all it remains are boundary conditions on the
gauge and Yukawa couplings (\ref{zoup-SOL5}), the $h=-MC$ relation,
and the soft scalar-mass sum rule (\ref{zoup-sumr}) at $M_{\rm GUT}$,
as applied in the two models.  Thus we examine the evolution of
these parameters according to their RGEs up
to two-loops for dimensionless parameters and at one-loop for
dimensionful ones with the relevant boundary conditions.  Below
$M_{\rm GUT}$ their evolution is assumed to be governed by the MSSM.
We further assume a unique supersymmetry breaking scale $M_{s}$ (which
we def\/ine as the geometric mean of the stop masses) and
therefore below that scale the  ef\/fective theory is just the SM.

\begin{figure}[t]
           \centerline{\includegraphics[width=10cm]{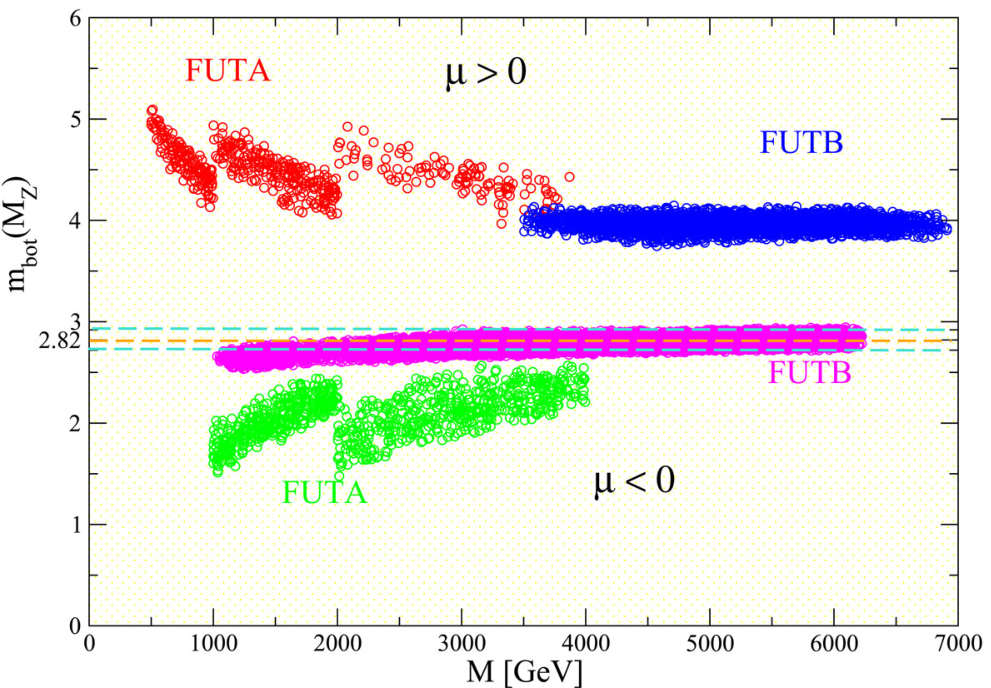}}
\vspace{5mm}
           \centerline{\includegraphics[width=10cm]{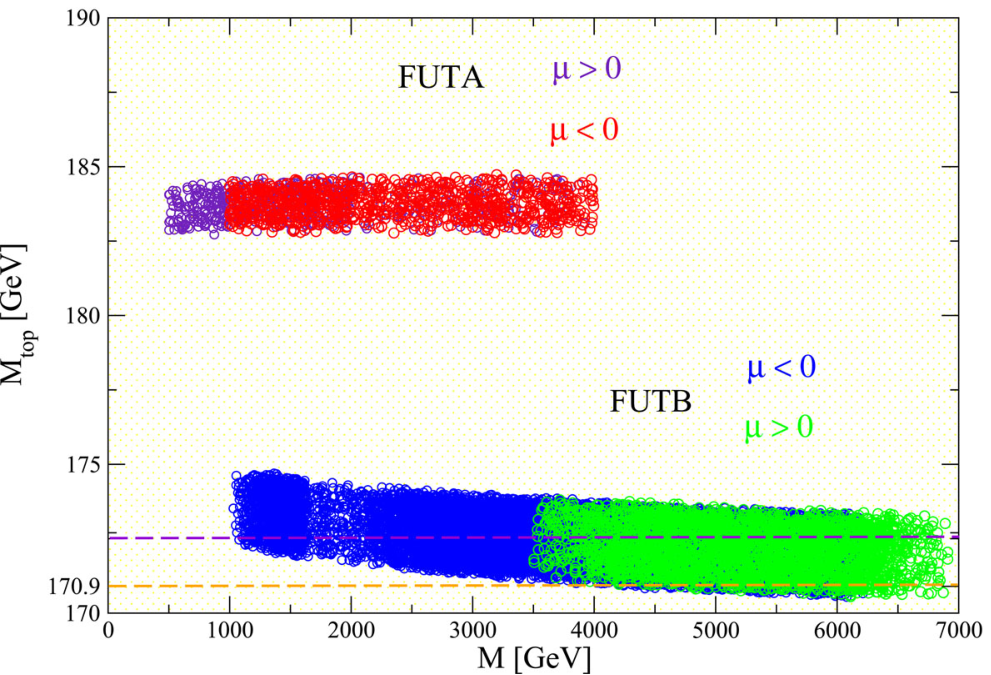}}

       \caption{The bottom quark mass at the $Z$~boson scale (upper)
                and top quark pole mass (lower plot) are shown
                as function of $M$ for both models.}
\label{fig:MtopbotvsM}
\end{figure}

We now present the comparison of the predictions of the four models with
the experimental data, see~\cite{thefullthing} for more details,
starting with the heavy quark masses.
In Fig.~\ref{fig:MtopbotvsM} we show the {\bf FUTA} and {\bf FUTB}
predictions for $M_{\rm top}$ and $m_{\rm bot}(M_Z)$ as a function of the
unif\/ied gaugi\-no mass~$M$, for the two cases
$\mu <0$ and $\mu >0$.
In the value of the bottom mass $m_{\rm bot}$,
we have included the corrections coming from bottom
squark-gluino loops and top squark-chargino loops~\cite{deltab}.  We
give the predictions for the running bottom quark mass evaluated at~$M_Z$, $m_{\rm bot}(M_Z) = 2.825 \pm  0.1$~\cite{jens}, to avoid the large
QCD uncertainties inherent for the pole mass.  The value of $m_{\rm bot}$ depends
strongly on the sign of $\mu$ due to the above mentioned
radiative corrections.  For both models~$\bf A$ and~$\bf B$ the values
for $\mu >0$ are above the central experimental value, with
$m_{\rm bot}(M_Z) \sim 4.0 - 5.0$~GeV.
For $\mu < 0$, on the other hand, model~$\bf B$ has
overlap with the experimental allowed values,
$m_{\rm bot}(M_Z) \sim 2.5-2.8$~GeV,
whereas for model~$\bf A$, $m_{\rm bot}(M_Z) \sim 1.5 - 2.6$~GeV, there is only a
small region of allowed parameter space at two sigma level, and only
for large values of $M$. This clearly selects the negative sign of~$\mu$.

The predictions for the top quark mass $M_{\rm top}$ are $\sim 183$ and
$\sim 172$~GeV in the models ${\bf A}$ and ${\bf B}$
respectively, as shown in the lower plot of Fig.~\ref{fig:MtopbotvsM}.
Comparing these predictions with the most recent experimental value
$M_{\rm top}^{\exp} = (170.9 \pm  1.8)$~GeV~\cite{mt1709}, and recalling
that the theoretical values for $M_{\rm top}$ may suf\/fer from a
correction of $\sim 4 \%$ \cite{zoup-acta}, we see that clearly model
${\bf B}$ is singled out.
In addition the value of $\tan \beta$ is found to be $\tan \beta \sim 54$ and
$\sim 48$ for models ${\bf A}$ and ${\bf B}$, respectively.
Thus from the comparison of the predictions of the two models with
experimental data only {\bf FUTB} with $\mu < 0$ survives.

We now analyze the impact of further low-energy observables on the model
{\bf FUTB} with $\mu < 0$. In the case where
all the soft scalar masses are universal at the unf\/ication scale,
there is no region of $M$ below ${\cal O}$(few~TeV) in which $m_{\tilde \tau} >
m_{\chi^0}$ is satisf\/ied (where $m_{\tilde \tau}$ is the lightest~$\tilde
\tau$ mass, and $m_{\chi^0}$ the lightest neutralino mass).  But once the
universality condition is relaxed this problem can be solved
naturally, thanks to the sum rule (\ref{zoup-sumr}).  Using this equation
and imposing the conditions of (a) successful radiative
electroweak symmetry breaking, (b) $m_{\tilde\tau}^2>0$ and (c)~$m_{\tilde\tau}> m_{\chi^0}$, a comfortable parameter space is found for
{\bf FUTB} with $\mu < 0$ (and also for {\bf FUTA} and both signs of $\mu$).

As  additional constraints we consider the following observables:
the rare $b$~decays $\br(b \to s \gamma)$ and $\br(B_s \to \mu^+ \mu^-)$,
the lightest Higgs boson mass
as well as the density of cold dark matter in the Universe, assuming it
consists mainly of neutralinos. More details and a complete set of
references can be found in~\cite{thefullthing}.

For the branching ratio $\br(b \to s \gamma)$, we take
the present
experimental value estimated by the Heavy Flavour Averaging
Group (HFAG) is~\cite{bsgexp}
\begin{gather*}
\br(b \to s \gamma ) = (3.55 \pm 0.24 {}^{+0.09}_{-0.10} \pm 0.03)
                       \times 10^{-4},
\end{gather*}
where the f\/irst error is the combined statistical and uncorrelated systematic
uncertainty, the latter two errors are correlated systematic theoretical
uncertainties and corrections respectively.

For the branching ratio $\br(B_s \to \mu^+ \mu^-)$, the SM prediction is
at the level of $10^{-9}$, while the present
experimental upper limit from the Tevatron is
$5.8 \times 10^{-8}$ at the $95\%$ C.L.~\cite{bsmmexp}, providing the
possibility for the MSSM to dominate the SM contribution.

Concerning the lightest Higgs boson mass, $M_h$, the SM bound of
$114.4$~GeV~\cite{LEPHiggsSM_MSSM} can be used. For the
prediction we use the code
{\tt FeynHiggs} \cite{feynhiggs,mhiggslong,mhiggsAEC,mhcMSSMlong}.

The lightest supersymmetric particle (LSP) is an
excellent candidate for cold dark matter (CDM)~\cite{EHNOS}, with a density
that falls naturally within the range
\begin{gather*}
0.094 < \Omega_{\rm CDM} h^2 < 0.129
\end{gather*}
favoured by a joint analysis of WMAP and other astrophysical and
cosmological data~\cite{WMAP}.

\begin{figure}[t]
           \centerline{\includegraphics[width=10cm]{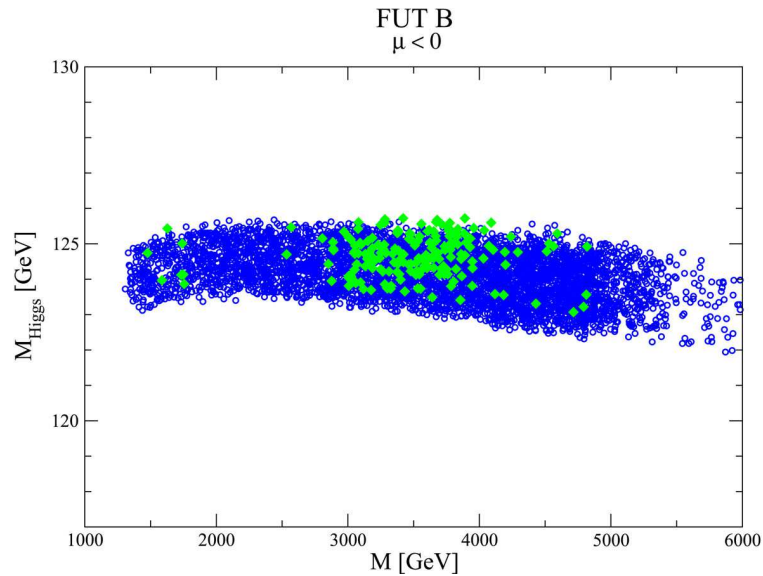}}
        \caption{The lightest Higgs mass, $M_h$,  as function of $M$ for
          the model {\bf FUTB} with $\mu < 0$, see text.}
\label{fig:Higgs}
\end{figure}

The prediction for $M_h$ of {\bf FUTB} with $\mu < 0$ is shown in
Fig.~\ref{fig:Higgs}.
The constraints from the two $B$~physics observables are taken into
account. In addition the CDM constraint (evaluated with
{\tt Micromegas}~\cite{micromegas}) is fulf\/illed for the lighter
(green) points in the plot, see~\cite{thefullthing} for details.
The lightest Higgs mass ranges in
\begin{gather}
M_h \sim 121-126~{\rm GeV} ,
\label{eq:Mhpred}
\end{gather}
where the uncertainty comes from
variations of the soft scalar masses, and
from f\/inite (i.e.~not logarithmically divergent) corrections in
changing renormalization scheme.  To this value one has to add $\pm  3$
GeV coming from unknown higher order corrections~\cite{mhiggsAEC}.
We have also included a small variation,
due to threshold corrections at the GUT scale, of up to $5 \%$ of the
FUT boundary conditions.
Thus, taking into account the $B$~physics constraints (and possibly the
CDM constraints) results naturally in a light Higgs boson that fulf\/ills
the LEP bounds~\cite{LEPHiggsSM_MSSM}.

In the same way the whole SUSY particle spectrum can be derived.
The resulting SUSY masses for {\bf FUTB} with $\mu < 0$ are rather large.
The lightest SUSY particle starts around 500~GeV, with the rest
of the spectrum being very heavy. The observation of SUSY particles at
the LHC or the ILC will only be possible in very favorable parts of the
parameter space. For most parameter combination only a SM-like light
Higgs boson in the range of equation~(\ref{eq:Mhpred}) can be observed.

We note that with such a heavy SUSY spectrum
the anomalous magnetic moment of the muon, \mbox{$(g-2)_\mu$}
(with $\amu \equiv (g-2)_\mu/2$), gives only a negligible correction to
the SM prediction.
The comparison of the experimental result and the SM value~\cite{DDDD}
\begin{gather*}
\amuexp-\amutheo = (27.5 \pm  8.4) \times 10^{-10}
\end{gather*}
would disfavor {\bf FUTB} with $\mu < 0$ by about $3 \sigma$. However,
since the SM is not regarded as excluded by $(g-2)_\mu$,
we still see {\bf FUTB} with $\mu < 0$ as the only surviving model.
A more detailed numerical analysis, also using
{\tt Suspect}~\cite{Djouadi:2002ze} for the RGE
running, and including all theory uncertainties for the models will be
presented in a future publication.

\section[Concluding remarks on the Realistic Finite Unified Theories]{Concluding remarks on the Realistic Finite Unif\/ied Theories}

The f\/initeness conditions in the supersymmetric part of the unbroken
theory lead to relations among the dimensionless couplings, i.e.\
{\it gauge-Yukawa unification}. In addition the f\/initeness conditions
in the SUSY-breaking sector of the theories lead to a
tremendous reduction of the number of the independent soft SUSY-breaking
parameters leaving one model ({\bf A}) with three and another ({\bf B})
with two free parameters.
Therefore the {\it finiteness-constrained MSSM}
consists of the well known MSSM with boundary conditions at the Grand
Unif\/ication scale for its va\-rious dimensionless and dimensionful
parameters inherited from the all-loop f\/initeness unbroken theories.
Obviously these lead to an extremely restricted and, consequently,
very predictive parameter space of the MSSM.

\section[Unified Theories from Fuzzy Higher Dimensions]{Unif\/ied Theories from Fuzzy Higher Dimensions}

Coset Space Dimensional Reduction (CSDR)
\cite{Forgacs:1979zs,Kapetanakis:hf,Kubyshin:vd,Bais:td,Farakos:1986sm,Chapline:wy,Barnes:ea,Hanlon,Manousselis:2001re} is a unif\/ication
scheme for obtaining realistic particle models from gauge theories on
higher D-dimensional spaces $M^D$. It suggests that a unif\/ication of
the gauge and Higgs sectors of the Standard Model can be achieved in
higher than four dimensions. Moreover the addition of fermions in the
higher-dimensional gauge theory leads naturally, after CSDR, to Yukawa
couplings in four dimensions. We present a study of the CSDR in the
non-commutative context which sets the rules for constructing new
particle models that might be phenomenologically interesting.  One
could study CSDR with the whole parent space $M^D$ being
non-commutative or with just non-commutative Minkowski space or
non-commutative internal space. We specialize here to this last
situation and therefore eventually we obtain Lorentz covariant
theories on commutative Minkowski space. We further specialize to
fuzzy non-commutativity, i.e.\ to matrix type non-commutativity. Thus,
following \cite{fuzzy1}, we consider non-commutative spaces like those
studied in~\cite{DV.M.K.,M.,Madore} and implementing the CSDR
principle on these spaces we obtain the rules for
constructing new particle models.

Next we reverse the above approach \cite{Aschieri:2006uw} and
examine how a four dimensional gauge theory dynamically develops
higher dimensions. The very concept of dimension
therefore gets an extra, richer dynamical perspective. We present a
simple f\/ield-theoretical model which realizes the above ideas. It is
def\/ined as a renormalizable $SU(N)$ gauge theory on four dimensional
Minkowski space $M^4$, containing 3 scalars in the adjoint of
$SU(N)$ that transform as vectors under an additional global $SO(3)$
symmetry with the most general renormalizable potential.  We then
show that the model dynamically develops fuzzy extra dimensions,
more precisely a fuzzy sphere $S^2_{N}$. The appropriate
interpretation is therefore as gauge theory on $M^4 \times S^2_{N}$.
The low-energy ef\/fective action is that of a four dimensional gauge
theory on $M^4$, whose gauge group and f\/ield content is dynamically
determined by compactif\/ication and dimensional reduction on the
internal sphere  $S^2_{N}$.  An interesting and  rich pattern of
spontaneous symmetry breaking appears, breaking the original $SU(N)$
gauge symmetry down to
 either $SU(n)$ or $SU(n_1)\times
SU(n_2) \times U(1)$. The latter case is the generic one, and
implies also a monopole f\/lux induced on the fuzzy sphere. The values
of $n_1$ and $n_2$ are determined dynamically.

We f\/ind moreover explicitly the tower of massive Kaluza--Klein modes
corresponding to the ef\/fective geometry, which justif\/ies the
interpretation as a compactif\/ied higher-dimensional gauge theory.
Nevertheless, the model is renormalizable.

A similar but dif\/ferent mechanism of dynamically generating extra
dimensions has been proposed some years ago in
\cite{Arkani-Hamed:2001ca}, known under the name of
``deconstruction''.  In this context, renormalizable four
dimensional asymptotically free gauge theories were considered,
which develop a ``lattice-like'' f\/ifth dimension. This idea
attracted considerable interest.  Our model is quite dif\/ferent, and
very simple: The $SU(N)$ gauge theory  is shown to develop fuzzy
extra dimensions through a standard symmetry breaking mechanism.

\section{The Fuzzy Sphere}

\subsection{Ordinary and Fuzzy spherical harmonics}\label{sec:sphere}

Let us start by recalling how to describe f\/ields on the 2-sphere.
The 2-sphere is a two-dimensional manifold embedded in $\mathbb{R}^3$,
with a global $SO(3) \sim SU(2)$ isometry group, def\/ined by the
equation
\begin{gather*}
x^2_1 + x^2_2 + x^2_3 = R^2
\end{gather*}
for a coordinate basis $x_{\hat{a}}$ in $\mathbb{R}^3$. We def\/ine the
coordinates $x_{\hat{a}}$ in terms of the spherical coordinates $y_a
= (\theta, \phi)$ and radius $R$ by,
\begin{gather*}
x_1  =  R \sin \theta \cos\phi, \qquad
x_2  =  R \sin \theta \sin \phi, \qquad
x_3  =  R \cos \theta,
\end{gather*}
which dictates the metric of the 2-sphere,
\begin{gather*}
ds^2 = R^2   d\theta^2 + R^2   \sin^2 \theta  d\phi^2.
\end{gather*}
The generators of $SU(2) \sim SO(3)$ are the angular momentum
operators $L_i$,
\begin{gather*}
L_{\hat{a}} = -i \varepsilon_{\hat{a}\hat{b}\hat{c}} x_{\hat{b}}
\partial_{\hat{c}}.
\end{gather*}
In terms of spherical coordinates the angular momentum operators are
\begin{gather*}
L_1  =    i   \sin \phi
\frac{\partial}{\partial \theta} + i \cos \phi \cot \theta
\frac{\partial}{\partial \phi}, \qquad
L_2  =  -i  \cos \phi  \frac{\partial}{\partial
\theta} + i\sin \phi  \cot \theta
\frac{\partial}{\partial \phi}, \qquad
L_3  =  -i  \frac{\partial}{\partial \phi},
\end{gather*}
which we can summarize as
\begin{gather*}
L_{\hat{a}} = -i k^{a}_{\hat{a}} \partial_a.
\end{gather*}
The metric tensor can also be expressed in terms of the Killing
vectors $k^{a}_{\hat{a}}$ (def\/ined by the above equations) as
\begin{gather*}
g^{ab} = \frac{1}{R^2}   k^{a}_{\hat{a}} k^b_{\hat{a}}.
\end{gather*}

We can expand any function on the 2-sphere in terms of the
eigenfunctions of the 2-sphere,
\begin{gather}\label{eq:expand}
a(\theta, \phi) = \sum^{\infty}_{l=0} \sum^l_{m=-l} a_{lm}
Y_{lm}(\theta, \phi),
\end{gather}
where $a_{lm}$ is a complex coef\/f\/icient and $Y_{lm}(\theta, \phi)$
are the spherical harmonics, which satisfy the equation
\begin{gather*}
L^2 Y_{lm} = -R^2 \Delta_{S^2} Y_{lm} = l(l+1) Y_{lm},
\end{gather*}
where $\Delta_{S^2}$ is the scalar Laplacian on the 2-sphere
\begin{gather*}
\Delta_{S^2} = \frac{1}{\sqrt{g}}   \partial_a ( g^{ab} \sqrt{g}
\partial_b).
\end{gather*}
The spherical harmonics have an eigenvalue $\mu \sim l(l+1)$ for
integer  $l = 0,1, \dots$, with degeneracy $2l+1$. The
orthogonality condition of the spherical harmonics is
\begin{gather*}
\int d\Omega \, Y^{\dag}_{lm} Y^{\phantom{\dag}}_{l'm'} = \delta_{l
l'} \delta_{m m'},
\end{gather*}
where $d\Omega = \sin \theta  d\theta d\phi$.

The spherical harmonics can be expressed in terms of the cartesian
coordinates $x_{\hat{a}}$ (with $\hat{a}=1,2,3$) of a unit vector in
$\mathbb{R}^{3}$,
\begin{gather}\label{eq:spharm}
Y_{lm}(\theta, \phi) = \sum_{\vec{a}} f ^{(lm)} _{\hat{a_{1}} \dots
\hat{a_{l}}} x^{\hat{a_{1}}}  \cdots x^{\hat{a_{l}}}
\end{gather}
where $f ^{(lm)} _{\hat{a_{1}} \dots \hat{a_{l}}}$ is a traceless
symmetric tensor of $SO(3)$ with rank $l$.

Similarly we can expand $N \times N$ matrices on a  sphere as,
\begin{gather}\label{eq:fuzzyexpand}
\hat{a}  =  \sum^{N-1}_{l=0} \sum^l_{m=-l} a_{lm} \hat{Y}_{lm}, \qquad
\hat{Y}_{lm}
 =  R^{-l}\sum_{\vec{a}} f ^{(lm)} _{\hat{a_{1}} \dots
\hat{a_{l}}} \hat{x}^{\hat{a_{1}}}  \cdots \hat{x}^{\hat{a_{l}}},
\end{gather}
where $\hat{x}_{\hat{a}}=\frac{2R}{\sqrt{N^{2}-1}}
X^{(N)}_{\hat{a}}$ are the generators of $SU(2)$ in the
$N$-dimensional representation and $f ^{(lm)}_{\hat{a_{1}} \dots
\hat{a_{l}}}$ is the same tensor as in (\ref{eq:spharm}). The
matrices $\hat{Y}_{lm}$ are known as fuzzy spherical harmonics for
reasons explained in the next subsection. They obey the
orthonormality condition
\begin{gather*}
\textrm{Tr}_N \left( \hat{Y}^{\dag}_{lm}
\hat{Y}_{l'm'}^{\phantom{\dag}} \right) = \delta_{l l'}   \delta_{m
m'}.
\end{gather*}
There is an obvious relation between equations (\ref{eq:expand}) and
(\ref{eq:fuzzyexpand}), namely
\begin{gather*}
\hat{a} = \sum^{N-1}_{l=0} \sum^{l}_{m=-l} a_{lm} \hat{Y}_{lm} \ \ \to  \ \
a(\theta, \phi) = \sum^{N-1}_{l = 0} \sum^l_{m = -l} a_{lm}
Y_{lm}(\theta, \phi).
\end{gather*}
Notice that the expansion in spherical harmonics is truncated at
$N-1$ ref\/lecting the f\/inite number of degrees of freedom in the
matrix $\hat{a}$. This allows the consistent def\/inition of a matrix
approximation of the sphere known as fuzzy sphere.

\subsection{The Matrix Geometry of the fuzzy sphere}

According to
the above discussion the fuzzy sphere \cite{Mad, Madore} is a matrix
approximation of the usual sphere $S^2$. The algebra of functions on
$S^2$ (for example spanned by the spherical harmonics) as explained
in the previous section is truncated at a given frequency and thus
becomes f\/inite dimensional. The truncation has to be consistent with
the associativity of the algebra and this can be nicely achieved
relaxing the commutativity property of the algebra. The fuzzy sphere
is the ``space'' described by this non-commutative algebra. The
algebra itself is that of $N\times N$ matrices. More precisely, the
algebra of functions on the ordinary sphere can be generated by the
coordinates of ${\mathbb{R}}^3$ modulo the relation $
\sum\limits_{\hat{a}=1}^{3} {x}_{\hat{a}}{x}_{\hat{a}} =r^{2}$. The fuzzy
sphere $S^2_{N}$ at fuzziness level $N-1$ is the non-commutative
manifold whose coordinate functions $i {X}_{\hat{a}}$ are $N \times
N$ hermitian matrices proportional to the generators of the
$N$-dimensional representation of $SU(2)$. They satisfy the
condition $ \sum\limits_{\hat{a}=1}^{3} X_{\hat{a}} X_{\hat{a}} = \alpha
r^{2}$ and the commutation relations
\begin{gather*}
[ X_{\hat{a}}, X_{\hat{b}} ] = C_{\hat{a} \hat{b} \hat{c}}
X_{\hat{c}},
\end{gather*}
where $C_{\hat{a} \hat{b} \hat{c}}= \varepsilon_{\hat{a} \hat{b}
\hat{c}}/r$ while the proportionality factor $\alpha$ goes as $N^2$
for $N$ large. Indeed it can be proven that for $N\to
\infty$ one obtains the usual commutative sphere.

On the fuzzy sphere there is a natural $SU(2)$ covariant
dif\/ferential calculus. This calculus is three-dimensional and the
derivations $e_{\hat{a}}$ along $X_{\hat{a}}$ of a function $ f$ are
given by $e_{\hat{a}}({f})=[X_{\hat{a}}, {f}]$. 
Accordingly the action of the Lie derivatives on functions is given
by
\begin{gather*}
{\cal L}_{\hat{a}} f = [{X}_{\hat{a}},f ];
\end{gather*}
these Lie derivatives satisfy the Leibniz rule and the $SU(2)$ Lie
algebra relation
\begin{gather*}
[ {\cal L}_{\hat{a}}, {\cal L}_{\hat{b}} ] = C_{\hat{a} \hat{b}
\hat{c}} {\cal L}_{\hat{c}}.
\end{gather*}
In the $N \to \infty$ limit the derivations $e_{\hat{a}}$
become $
e_{\hat{a}} = C_{\hat{a} \hat{b} \hat{c}} x^{\hat{b}}
\partial^{\hat{c}}
$ and only in this commutative limit the tangent space becomes
two-dimensional. The exterior derivative is given by
\begin{gather*}
d f = [X_{\hat{a}},f]\theta^{\hat{a}}
\end{gather*}
with $\theta^{\hat{a}}$ the one-forms dual to the vector f\/ields
$e_{\hat{a}}$,
$\langle e_{\hat{a}},\theta^{\hat{b}}\rangle =\delta_{\hat{a}}^{\hat{b}}$. The
space of one-forms is generated by the $\theta^{\hat{a}}$'s in the
sense that for any one-form $\omega=\sum\limits_i f_i d h_i  t_i$ we can
always write
$\omega=\sum\limits_{\hat{a}=1}^3{\omega}_{\hat{a}}\theta^{\hat{a}}$ with
given functions $\omega_{\hat{a}}$ depending on the functions $f_i$,
$h_i$ and $t_i$. The action of the Lie derivatives ${\cal
L}_{\hat{a}}$ on the one-forms $\theta^{\hat{b}}$ explicitly reads
\begin{gather*}
{\cal L}_{\hat{a}}(\theta^{\hat{b}}) =  C_{\hat{a}\hat{b}\hat{c}}
\theta^{\hat{c}}.
\end{gather*}
On a general one-form $\omega=\omega_{\hat{a}}\theta^{\hat{a}}$ we
have $ {\cal L}_{\hat{b}}\omega={\cal
L}_{\hat{b}}(\omega_{\hat{a}}\theta^{\hat{a}})=
\left[X_{\hat{b}},\omega_{\hat{a}}\right]\theta^{\hat{a}}-\omega_{\hat{a}}C^{\hat{a}}_{\
\hat{b} \hat{c}}\theta^{\hat{c}} $ and therefore
\begin{gather*}
({\cal
L}_{\hat{b}}\omega)_{\hat{a}}=\left[X_{\hat{b}},\omega_{\hat{a}}\right]-
\omega_{\hat{c}}C^{\hat{c}}_{\ \hat{b}  \hat{a}};
\end{gather*}
this formula will be fundamental for formulating the CSDR principle
on fuzzy cosets.

The dif\/ferential geometry on  the product space Minkowski times
fuzzy sphere, $M^{4} \times  S^2_{N}$, is easily obtained from that
on $M^4$ and on $S^2_N$. For example a one-form $A$ def\/ined on
$M^{4} \times  S^2_{N}$ is written as{\samepage
\begin{gather*}
A= A_{\mu} dx^{\mu} + A_{\hat{a}} \theta^{\hat{a}}
\end{gather*}
with $A_{\mu} =A_{\mu}(x^{\mu}, X_{\hat{a}} )$ and $A_{\hat{a}}
=A_{\hat{a}}(x^{\mu}, X_{\hat{a}} )$.}

One can also introduce spinors on the fuzzy sphere and study the Lie
derivative on these spinors. Although here we have sketched the
dif\/ferential geometry on the fuzzy sphere,  one can study other
(higher-dimensional) fuzzy spaces (e.g.~fuzzy $CP^M$) and with
similar techniques their dif\/ferential geometry.

\section{Dimensional Reduction of Fuzzy Extra Dimensions}

\subsection[Actions in higher dimensions seen as four-dimensional
actions (expansion in Kaluza-Klein modes)]{Actions in higher dimensions seen as four-dimensional
actions\\ (expansion in Kaluza--Klein modes)}

 First we consider on
$M^{4} \times  (S/R)_{F}$ a non-commutative gauge theory with gauge
group $G=U(P)$ and examine its four-dimensional interpretation.
$(S/R)_{F}$ is a fuzzy coset, for example the fuzzy sphere
$S^{2}_{N}$. The action is
\begin{gather}\label{formula8}
{\cal A}_{YM}=\frac{1}{4g^{2}} \int d^{4}x k\,{\rm Tr}\, {\rm tr}_{G}
F_{MN}F^{MN},
\end{gather}
where $kTr$ denotes integration over the fuzzy coset $(S/R)_F $
described by $N\times  N$ matrices; here the parameter $k$ is related
to the size of the fuzzy coset space. For example for the fuzzy
sphere we have $r^{2} = \sqrt{N^{2}-1}\pi k$~\cite{Madore}. In the
$N\to \infty$ limit $k\,{\rm Tr}$ becomes the usual integral on the
coset space. For f\/inite $N$, ${\rm Tr}$ is a good integral because it has
the cyclic property ${\rm Tr}(f_1\cdots f_{p-1}f_p)={\rm Tr}(f_pf_1\cdots
f_{p-1})$. It is also invariant under the action of the group $S$,
that is  inf\/initesimally given by the Lie derivative. In the action
(\ref{formula8}) ${\rm tr}_G$ is the gauge group $G$ trace. The
higher-dimensional f\/ield strength $F_{MN}$, decomposed in
four-dimensional space-time and extra-dimensional components, reads
as follows $(F_{\mu \nu}, F_{\mu \hat{b}}, F_{\hat{a} \hat{b} })$;
explicitly the various components of the f\/ield strength are given by
\begin{gather*}
F_{\mu \nu} =
\partial_{\mu}A_{\nu} -
\partial_{\nu}A_{\mu} + [A_{\mu}, A_{\nu}],\\
F_{\mu \hat{a}} =
\partial_{\mu}A_{\hat{a}} - [X_{\hat{a}}, A_{\mu}] + [A_{\mu},
A_{\hat{a}}], \nonumber\\
F_{\hat{a} \hat{b}} =   [ X_{\hat{a}}, A_{\hat{b}}] - [
X_{\hat{b}}, A_{\hat{a}} ] + [A_{\hat{a}} , A_{\hat{b}} ] -
C^{\hat{c}}_{\ \hat{a} \hat{b}}A_{\hat{c}}.
\end{gather*}
Under an inf\/initesimal $ G $ gauge transformation
$\lambda=\lambda(x^{\mu},X^{\hat{a}})$ we have
\begin{gather*}
\delta A_{\hat{a}} = -[ X_{\hat{a}}, \lambda] +
[\lambda,A_{\hat{a}}],
\end{gather*}
thus $F_{MN}$ is covariant under {local} $G$ gauge transformations:
$F_{MN}\to F_{MN}+[\lambda, F_{MN}]$. This is an
inf\/initesimal Abelian $U(1)$ gauge transformation if $\lambda$ is
just an antihermitian function of the coordinates $x^\mu$,
$X^{\hat{a}}$ while it is an inf\/initesimal non-Abelian $U(P)$ gauge
transformation if $\lambda$ is valued in ${\rm{Lie}}(U(P))$, the Lie
algebra of hermitian $P\times  P$ matrices. In the following we will
always assume ${\rm{Lie}}(U(P))$ elements to commute with the
coordinates $X^{\hat{a}}$. In fuzzy/non-commutative gauge theory and
in Fuzzy-CSDR a fundamental role is played by the covariant
coordinate,
\begin{gather*}
\varphi_{\hat{a}} \equiv X_{\hat{a}} + A_{\hat{a}}.
\end{gather*}
This f\/ield transforms indeed covariantly under a gauge
transformation, $
\delta(\varphi_{\hat{a}})=[\lambda,\varphi_{\hat{a}}]~. $ In terms
of $\varphi$ the f\/ield strength in the non-commutative directions
reads,
\begin{gather}
F_{\mu \hat{a}} =
\partial_{\mu}\varphi_{\hat{a}} + [A_{\mu}, \varphi_{\hat{a}}]=
D_{\mu}\varphi_{\hat{a}},\qquad
F_{\hat{a} \hat{b}} = [\varphi_{\hat{a}}, \varphi_{\hat{b}}] -
C^{\hat{c}}_{\ \hat{a} \hat{b}} \varphi_{\hat{c}};
\label{action-CSDR}
\end{gather}
and using these expressions the action reads
\begin{gather}
{\cal A}_{YM}= \int d^{4}x\, {\rm Tr}\, {\rm tr}_{G} \left( \frac{k}
{4g^{2}}F_{\mu \nu}^{2} + \frac{k}{2g^{2}}(D_{\mu}\varphi_{\hat{a}})^{2} -
V(\varphi)\right),\label{theYMaction}
\end{gather}
where the potential term $V(\varphi)$ is the $F_{\hat{a} \hat{b}}$
kinetic term (in our conventions $F_{\hat{a} \hat{b}}$ is
antihermitian so that $V(\varphi)$ is hermitian and non-negative)
\begin{gather}
V(\varphi)=-\frac{k}{4g^{2}} \,{\rm Tr}\,{\rm tr}_G \sum_{\hat{a} \hat{b}}
F_{\hat{a} \hat{b}} F_{\hat{a} \hat{b}}
\nonumber \\
\phantom{V(\varphi)=}{} =-\frac{k}{4g^{2}} \,{\rm Tr}\,{\rm tr}_G \left( [\varphi_{\hat{a}},
\varphi_{\hat{b}}][\varphi^{\hat{a}}, \varphi^{\hat{b}}] -
4C_{\hat{a} \hat{b} \hat{c}} \varphi^{\hat{a}} \varphi^{\hat{b}}
\varphi^{\hat{c}} + 2r^{-2}\varphi^{2} \right).\label{pot1}
\end{gather}
The action (\ref{theYMaction}) is naturally interpreted as an action
in four dimensions. The inf\/initesimal $G$ gauge transformation with
gauge parameter $\lambda(x^{\mu},X^{\hat{a}})$ can indeed be
interpreted just as an $M^4$ gauge transformation. We write
\begin{gather}
\lambda(x^{\mu},X^{\hat{a}})=\lambda^{\alpha}(x^{\mu},X^{\hat{a}}){\cal
T}^{\alpha} =\lambda^{h, \alpha}(x^{\mu})T^{h}{\cal
T}^{\alpha},\label{3.33}
\end{gather}
where ${\cal T}^{\alpha}$ are hermitian generators of $U(P)$,
$\lambda^{\alpha}(x^\mu,X^{\hat{a}})$ are $n\times  n$ antihermitian
matrices and thus are expressible as $\lambda(x^\mu)^{\alpha ,
h}T^{h}$, where $T^{h}$ are antihermitian generators of $U(n)$. The
f\/ields $\lambda(x^{\mu})^{\alpha , h}$, with $h=1,\ldots, n^2$, are
the Kaluza--Klein modes of $\lambda(x^{\mu}, X^{\hat{a}})^{\alpha}$.
We now consider on equal footing the indices $h$ and $\alpha$ and
interpret the f\/ields on the r.h.s.\ of~(\ref{3.33}) as one f\/ield
valued in the tensor product Lie algebra ${\rm{Lie}}(U(n)) \otimes
{\rm{Lie}}(U(P))$. This Lie algebra is indeed ${\rm{Lie}}(U(nP))$
(the $(nP)^2$ generators $T^{h}{\cal T}^{\alpha}$ being $nP\times
nP$ antihermitian matrices that are linear independent). Similarly
we rewrite the gauge f\/ield $A_\nu$ as
\begin{gather*}
A_\nu(x^{\mu},X^{\hat{a}})=A_{\nu}^{\alpha}(x^{\mu},X^{\hat{a}}){\cal
T}^{\alpha} =A_{\nu}^{h, \alpha}(x^{\mu})T^{h}{\cal T}^{\alpha},
\end{gather*}
and interpret it as a ${\rm{Lie}}(U(nP))$ valued gauge f\/ield on
$M^4$, and similarly for $\varphi_{\hat{a}}$. Finally ${\rm Tr}\, {\rm tr}_{G}$
is the trace over $U(nP)$ matrices in the fundamental
representation.

Up to now we have just performed a ordinary fuzzy dimensional
reduction. Indeed in the commutative case the expression
(\ref{theYMaction}) corresponds to rewriting the initial lagrangian
on $M^4\times  S^2$ using spherical harmonics on $S^2$. Here the
space of functions is f\/inite dimensional and therefore the inf\/inite
tower of modes reduces to the f\/inite sum given by~${\rm Tr}$.

\subsection{Non-trivial Dimensional reduction in the case of Fuzzy
Extra Dimensions}

Next we  reduce the number of gauge f\/ields and
scalars in the action (\ref{theYMaction}) by applying the Coset
Space Dimensional Reduction (CSDR) scheme. Since $SU(2)$ acts on the
fuzzy sphere $(SU(2)/U(1))_F$, and more in general  the group $S$
acts on the fuzzy coset $(S/R)_F$, we can state the CSDR principle
in the same way as in the continuum case, i.e.\ the f\/ields in the
theory must be invariant under the inf\/initesimal $SU(2)$,
respectively $S$, action up to an inf\/initesimal gauge transformation
\begin{gather*}
{\cal L}_{\hat{b}} \phi = \delta_{W_{\hat{b}}}\phi= W_{\hat{b}}\phi,
\qquad
{\cal L}_{\hat{b}}A = \delta_{W_{\hat{b}}}A=-DW_{\hat{b}},
\end{gather*}
where $A$ is the one-form gauge potential $A = A_{\mu}dx^{\mu} +
A_{\hat{a}} \theta^{\hat{a}}$, and $W_{\hat{b}}$ depends only on the
coset coordinates $X^{\hat{a}}$ and (like $A_\mu, A_a$) is
antihermitian. We thus write $W_{\hat{b}}=W_{\hat{b}}^{\alpha}{\cal
T}^{\alpha}$, $\alpha=1,2,\ldots, P^2,$ where ${\cal  T}^i$ are
hermitian generators of $U(P)$ and $(W_b^i)^\dagger=-W_b^i$, here
${}^\dagger$ is hermitian conjugation on the $X^{\hat{a}}$'s.

In terms of the covariant coordinate $\varphi_{\hat{d}} =X_{\hat{d}}
+ A_{\hat{d}}$ and of
\begin{gather*}
\omega_{\hat{a}} \equiv X_{\hat{a}} - W_{\hat{a}},
\end{gather*}
the CSDR constraints assume a particularly simple form, namely
\begin{gather}\label{3.19}
[\omega_{\hat{b}}, A_{\mu}] =0,
\\
\label{eq7}
C_{\hat{b} \hat{d} \hat{e}} \varphi^{\hat{e}} = [\omega_{\hat{b}},
\varphi_{\hat{d}} ].
\end{gather}
In addition we  have a consistency condition  following from the
relation $[{\cal{L}}_{\hat{a}},{\cal{L}}_{\hat{b}}]=
C_{\hat{a}\hat{b}}^{~~\hat{c}}{\cal{L}}_{\hat{c}}$:
\begin{gather}\label{3.17}
[ \omega_{\hat{a}} , \omega_{\hat{b}}] = C_{\hat{a} \hat{b}}^{\ \
\hat{c}} \omega_{c},
\end{gather}
where $\omega_{\hat{a}}$ transforms as $ \omega_{\hat{a}}\to
\omega'_{\hat{a}} = g\omega_{\hat{a}}g^{-1}. $ One proceeds in a
similar way for the spinor f\/ields~\cite{fuzzy1}.

\subsubsection{Solving the CSDR constraints for
the fuzzy sphere}

We consider $(S/R)_{F}=S^{2}_{N}$, i.e.\ the fuzzy
sphere, and to be def\/inite at fuzziness level $N-1$ ($N \times  N$
matrices). We study here the basic example where the gauge group is
$G=U(1)$. In this case the
$\omega_{\hat{a}}=\omega_{\hat{a}}(X^{\hat{b}})$ appearing in the
consistency condition (\ref{3.17}) are $N \times  N$ antihermitian
matrices and therefore can be interpreted as elements of
${\rm{Lie}}(U(N))$. On the other hand the $\omega_{\hat{a}}$ satisfy
the commutation relations (\ref{3.17}) of ${\rm{Lie}}(SU(2))$.
Therefore in order to satisfy the consistency condition (\ref{3.17})
we have to embed ${\rm{Lie}}(SU(2))$ in ${\rm{Lie}}(U(N))$. Let
$T^h$ with $h = 1, \ldots ,(N)^{2}$ be the generators of
${\rm{Lie}}(U(N))$ in the fundamental representation, we can always
use the convention $h= (\hat{a} , u)$ with $\hat{a} = 1,2,3$ and $u=
4,5,\ldots, N^{2}$ where the $T^{\hat{a}}$ satisfy the $SU(2)$ Lie
algebra,
\begin{gather*}
[T^{\hat{a}}, T^{\hat{b}}] = C^{\hat{a} \hat{b}}_{\ \
\hat{c}}T^{\hat{c}}.
\end{gather*}
Then we def\/ine an embedding by identifying
\begin{gather}
 \omega_{\hat{a}}= T_{\hat{a}}.
\label{embedding}
\end{gather}
The constraint (\ref{3.19}), $[\omega_{\hat{b}} , A_{\mu}] = 0$,
then implies that the four-dimensional gauge group $K$ is the
centralizer of the image of $SU(2)$ in $U(N)$, i.e.\
\[
K=C_{U(N)}(SU((2))) = SU(N-2) \times  U(1)\times  U(1),
\]  where the
last $U(1)$ is the $U(1)$ of $U(N)\simeq SU(N)\times  U(1)$. The
functions $A_{\mu}(x,X)$ are arbitrary functions of $x$ but the $X$
dependence is such that $A_{\mu}(x,X)$ is ${\rm{Lie}}(K)$ valued
instead of ${\rm{Lie}}(U(N))$, i.e.\ eventually we have a
four-dimensional gauge potential $A_\mu(x)$ with values in
${\rm{Lie}}(K)$. Concerning the constraint~(\ref{eq7}), it is
satisf\/ied by choosing
\begin{gather}
\label{soleasy} \varphi_{\hat{a}}=r \varphi(x) \omega_{\hat{a}}~,
\end{gather}
i.e.\ the unconstrained degrees of freedom correspond to the scalar
f\/ield $\varphi(x)$ which is a singlet under the four-dimensional
gauge group~$K$.

The choice (\ref{embedding}) def\/ines one of the possible embedding
of ${\rm{Lie}}(SU(2))$ in ${\rm{Lie}}(U(N))$. For example, we could
also embed ${\rm{Lie}}(SU(2))$ in ${\rm{Lie}}(U(N))$ using the
irreducible $N$-dimensional rep. of $SU(2)$, i.e.\ we could identify
$\omega_{\hat{a}}= X_{\hat{a}}$. The constraint~(\ref{3.19}) in this
case implies that the four-dimensional gauge group is $U(1)$ so that
$A_\mu(x)$ is $U(1)$ valued. The constraint~(\ref{eq7}) leads again
to the scalar singlet $\varphi(x)$.

In general, we start with a $U(1)$ gauge theory on $M^4\times
S^2_N$. We solve the CSDR constraint~(\ref{3.17}) by embedding
$SU(2)$ in $U(N)$. There exist $p_{N}$ embeddings, where $p_N$ is
the number of ways one can partition the integer $N$ into a set of
non-increasing positive integers \cite{Mad}. Then the constraint
(\ref{3.19}) gives the surviving four-dimensional gauge group. The
constraint (\ref{eq7}) gives the surviving four-dimensional scalars
and equation (\ref{soleasy}) is always a solution but in general not the
only one. By setting $\phi_{\hat{a}}=\omega_{\hat{a}}$ we obtain
always a minimum of the potential. This minimum is given by the
chosen embedding of $SU(2)$ in $U(N)$.

An important point that we would like to stress here is the question
of the renormalizability of the gauge theory def\/ined on $M_4 \times
(S/R)_F$. First we notice that the theory exhibits certain features
so similar to a higher-dimensional gauge theory def\/ined on $M_4
\times  S/R$ that naturally it could be considered as a
higher-dimensional theory too. For instance the isometries of the
spaces $M_4 \times  S/R$ and $M_4 \times  (S/R)_F$ are the same. It
does not matter if the compact space is fuzzy or not. For example in
the case of the fuzzy sphere, i.e.\ $M_4 \times  S^2_N$, the
isometries are $SO(3,1) \times  SO(3)$ as in the case of the
continuous space, $M_4 \times  S^2$. Similarly the coupling of a~gauge theory def\/ined on $M_4 \times  S/R$ and on $M_4 \times  (S/R)_F$
are both dimensionful and have exactly the same dimensionality. On
the other hand the f\/irst theory is clearly non-renormalizable, while
the latter is renormalizable (in the sense that divergencies can be
removed by a f\/inite number of counterterms). So from this point of
view one f\/inds a partial justif\/ication of the old hopes for
considering quantum f\/ield theories on non-commutative structures. If
this observation can lead  to f\/inite theories too, it remains as an
open question.

\section{Dynamical Generation of Extra Dimensions}

Let us now discuss a further development \cite{Aschieri:2006uw} of
these ideas,
 which addresses in detail the questions of
 quantization and renormalization. This leads to a slightly
modif\/ied model with an extra term in the potential, which
dynamically selects a unique (nontrivial) vacuum out of the many
possible CSDR solutions, and moreover generates a magnetic f\/lux on
the fuzzy sphere. It also allows to show that the full tower of
Kaluza--Klein modes is generated on $S^2_N$.

\subsection{The four dimensional action}

We start with a $SU(N)$ gauge theory on four dimensional Minkowski
space $M^4$ with coordinates~$y^\mu$, $\mu = 0,1,2,3$.  The action
under consideration is
\begin{gather*}
{\cal S}_{YM}= \int d^{4}y\, {\rm Tr}\left( \frac{1}{4g^{2}}\, F_{\mu
\nu}^\dagger F_{\mu \nu} + (D_{\mu}\phi_{{a}})^\dagger
D_{\mu}\phi_{{a}}\right) - V(\phi), 
\end{gather*}
where $A_\mu$ are $SU(N)$-valued gauge f\/ields, $D_\mu =
\partial_\mu + [A_\mu,\cdot]$, and
\[
 \phi_{{a}} = - \phi_{{a}}^\dagger , \qquad a=1,2,3
 \] are 3
antihermitian scalars in the adjoint of $SU(N)$,
\[
 \phi_{{a}} \to
U^\dagger \phi_{{a}} U,
\]
where $U = U(y) \in SU(N)$. Furthermore,
the $\phi_a$ transform as vectors of an additional global $SO(3)$
symmetry. The potential $V(\phi)$ is taken to be the most general
renormalizable action invariant under the above symmetries, which is
\begin{gather}
V(\phi) = {\rm Tr}  \left( g_1 \phi_a\phi_a \phi_b\phi_b +
g_2\phi_a\phi_b\phi_a \phi_b - g_3 \varepsilon_{a b c} \phi_a \phi_b
\phi_c + g_4\phi_a \phi_a \right) \nonumber\\
\phantom{V(\phi) =}{}  + \frac{g_5}{N}\,
{\rm Tr}\, (\phi_a \phi_a)\,{\rm Tr}\,(\phi_b \phi_b) + \frac{g_6}{N} \,{\rm Tr}\, (\phi_a
\phi_b)\,{\rm Tr}\, (\phi_a \phi_b) +g_7. \label{pot}
\end{gather}
This may not look very transparent at f\/irst sight, however it can be
written in a very intuitive way. First, we make the scalars
dimensionless by rescaling
\[ \phi'_a = R  \phi_a,
\] where $R$ has
dimension of length; we will usually suppress $R$ since it can
immediately be reinserted, and drop the prime from now on.  Now
observe that for a suitable choice of $R$,
\[
R = \frac{2 g_2}{g_3}, 
\] the potential can be rewritten as
\begin{gather*}
 V(\phi)= {\rm Tr} \left( a^2
(\phi_a\phi_a + \tilde b  \one)^2 + c +\frac{1}{\tilde g^2}
F_{ab}^\dagger F_{ab}  \right) + \frac{h}{N}  g_{ab} g_{ab}
\end{gather*}
for suitable constants $a$, $b$, $c$, $\tilde g$, $h$, where
\begin{gather*}
F_{{a}{b}} =  [\phi_{{a}}, \phi_{{b}}] -
\varepsilon_{abc} \phi_{{c}}  = \varepsilon_{abc} F_c , \nonumber\\
\tilde b  =  b + \frac{d}{N} \, {\rm Tr}\,(\phi_a \phi_a), \qquad
g_{ab} = {\rm Tr}\,(\phi_a \phi_b). 
\end{gather*}
We will omit $c$ from now.
Notice that two couplings were reabsorbed in the def\/initions of $R$ and~$\tilde b$.
 The potential is clearly positive
def\/inite provided
\[
 a^2 = g_1+g_2 >0, \qquad \frac 2{\tilde g^2} =
- g_2 >0, \qquad h \geq 0,
\] which we assume from now on.  Here
$\tilde b = \tilde b(y)$ is a scalar, $g_{ab} = g_{ab}(y)$ is a
symmetric tensor under the global $SO(3)$, and $F_{ab}=F_{ab}(y)$ is
a $su(N)$-valued antisymmetric tensor f\/ield which will be
interpreted as f\/ield strength in some dynamically generated extra
dimensions below.  In this form, $V(\phi)$ looks like the action of
Yang--Mills gauge theory on a fuzzy sphere in the matrix formulation
\cite{Steinacker:2003sd,Steinacker:2004yu,Carow-Watamura:1998jn,
Presnajder:2003ak}. It dif\/fers from the potential in
(\ref{action-CSDR}) only by the presence of the f\/irst term $a^2
(\phi_a\phi_a + \tilde b)^2$, which is strongly suggested
 by renormalization.
In fact it is necessary for the interpretation as pure YM action,
and we will see that it is very welcome on physical grounds since it
dynamically determines and stabilizes a vacuum, which can be
interpreted as extra-dimensional fuzzy sphere. In particular, it
removes unwanted f\/lat directions.

\subsection{Emergence of extra dimensions and the fuzzy sphere}
\label{sec:emergence}

The vacuum of the above model is given by the minimum of the potential
(\ref{pot}). Finding the minimum of the potential is a rather
nontrivial task, and the answer depends crucially on the parameters in
the potential \cite{Aschieri:2006uw}. The conditions for the global
minimum imply that $\phi_a$ is a~representation of $SU(2)$, with Casimir
$\tilde b$ (where it was assumed for simplicity $ h = 0$).  Then, it
is easy to write down a large class of solutions to the minimum of the
potential, by noting that any decomposition of $N = n_1 N_1 + \cdots +
n_h N_h$ into irreps of $SU(2)$ with multiplicities $n_i$ leads to a
block-diagonal solution
\begin{gather} \phi_a = {\rm diag}\,\big(\a_1\, X_a^{(N_1)}, \dots,
\a_k\, X_a^{(N_k)}\big)
\label{solution-general}
\end{gather}
of the vacuum equations, where $\a_i$ are suitable
constants which will be determined below.

It turns out \cite{Aschieri:2006uw} that
there are essentially only 2 types of vacua:
\begin{enumerate}\itemsep=0pt
\item {\em Type I vacuum.}
 It is plausible that the solution (\ref{solution-general}) with minimal
potential contains only representations whose Casimirs are close to
$\tilde b$. In particular, let $M$ be the dimension of the irrep whose
Casimir $C_2(M)\approx \tilde b$ is closest to $\tilde b$. If
furthermore the dimensions match as $N = M n$, we expect that the
vacuum is given by $n$ copies of the irrep $(M)$, which can be written
as
$\phi_a = \alpha  X_a^{(N)} \otimes 1_{n}$ with low-energy
gauge group $SU(n)$.
\item {\em Type II vacuum.}
Consider again a solution (\ref{solution-general}) with $n_i$ blocks of
size $N_i = \tilde N +m_i$, where $\tilde N$ is def\/ined by
$\tilde b = \frac 14 (\tilde N^2-1)$, and assume that $\tilde N$ is large and
$\frac{m_i}{\tilde N} \ll 1$.  The action
is then given by
\begin{gather*}
 V(\phi) = {\rm Tr} \left( \frac {1}{2\tilde g^2}
\sum_i n_i\, m_i^2   \one_{N_i}   + O\left(\frac 1{N_i}\right) \right) \approx
\frac{1}{2\tilde{g}^{2}}  \frac{N}{k}  \sum_{i} n_i   m_{i}^{2} ,
\end{gather*}
 where $k=\sum n_i$ is the total number of irreps, and the solution
can be interpreted in terms of ``instantons'' (non-Abelian monopoles)
on the internal fuzzy sphere \cite{Steinacker:2003sd}.  Hence in order
to determine the solution of type (\ref{solution-general}) with minimal
action, we simply have to minimize $\sum_i n_i   m_i^2$, where the $m_i
\in \Z -\tilde N$ satisfy the constraint $\sum n_i  m_i = N - k \tilde
N$.  In this case the solution with minimal potential among
all possible partitions (\ref{solution-general}) is given by
\[
 \phi_a = \left(\begin{array}{cc}\alpha_1  X_a^{(N_1)}\otimes
1_{n_1} & 0 \\ 0 & \alpha_2\,X_a^{(N_2)}\otimes 1_{n_2}
             \end{array}\right) ,
\]
with low-energy gauge group $SU(n_1)\times  SU(n_2) \times  U(1)$.
\end{enumerate}

Again, the $X_a^{(N)}$ are interpreted as coordinate functions of a
fuzzy sphere $S^2_{N}$, and the ``scalar'' action
\[
 S_{\phi} = {\rm Tr}\, V(\phi) = {\rm Tr}  \left(a^2 (\phi_a\phi_a + \tilde b)^2 +
\frac 1{\tilde g^2}  F_{ab}^\dagger F_{ab}\right) 
\] for
$N \times  N$ matrices $\phi_a$ is precisely the action for a $U(n)$
Yang--Mills theory on $S^2_{N}$ with coupling~$\tilde g$, as shown in
\cite{Steinacker:2003sd}. In fact, the new term $(\phi_a\phi_a +
\tilde b)^2$ is essential for this interpretation, since it
stabilizes the vacuum $\phi_a = X_a^{(N)}$ and gives a large mass to
the extra ``radial'' scalar f\/ield which otherwise arises.  The
f\/luctuations of $\phi_a = X_a^{(N)} + A_a$ then provide the
components $A_a$ of a~higher-dimensional gauge f\/ield $A_M = (A_\mu,
A_a)$, and the action can be interpreted as YM theory on the
6-dimensional space $M^4 \times  S^2_{N}$, with gauge group depending
on the particular vacuum.  We therefore interpret the vacuum as
describing dynamically generated extra dimensions in the form of a
fuzzy sphere $S^2_{N}$. This geometrical interpretation can be fully
justif\/ied
 by working out the spectrum of Kaluza--Klein
modes.  The ef\/fective low-energy theory is then given by the zero
modes on $S^2_{N}$. This approach provides a clear dynamical
selection of the geometry due to the term $(\phi_a\phi_a + \tilde
b)^2$ in the action.

Perhaps the most remarkable aspect of this model is that the geometric
interpretation and the corresponding low-energy degrees of freedom
depend in a nontrivial way on the parameters of the model, which are
running under the RG group. Therefore the massless degrees of freedom
and their geometrical interpretation depend on the energy scale. In
particular, the low-energy gauge group generically turns out to be
$SU(n_1) \times  SU(n_2)\times  U(1)$ or $SU(n)$, while gauge groups which are
products of more than two simple components (apart from $U(1)$) do not
seem to occur. The values of $n_1$ and $n_2$ are determined
dynamically, and with the appropriate choice of parameters it is
possible to construct vacuum solutions where they are as small,  such
as~2 and~3~\cite{Aschieri:2006uw}.

It is interesting to examine  the running of the coupling constants
under the RG. $R$ turns out to run only logarithmically, implies
that the scale of the internal spheres is only mildly af\/fected by
the RG f\/low. However, $\tilde b$ is running essentially
quadratically, hence is generically large. This is quite welcome
here: starting with some large $N$, $\tilde{b} \approx
C_2(\tilde{N})$ must indeed be large in order to lead to the
geometric interpretation discussed above. Hence the problems of
naturalness or f\/ine-tuning appear to be rather mild here.

A somewhat similar model has been studied  in
\cite{Andrews:2005cv,Andrews:2006aw}, which realizes deconstruction
and a ``twisted'' compactif\/ication of an extra fuzzy sphere based on
a supersymmetric gauge theory. Our model is dif\/ferent and does not
require supersymmetry, leading to a much richer pattern of symmetry
breaking and ef\/fective geometry. For other relevant work see e.g.~\cite{M.}.

The dynamical formation of fuzzy spaces found here is also related
to recent work studying the emergence of stable submanifolds in
modif\/ied IIB matrix models. In particular, previous studies based on
actions for fuzzy gauge theory dif\/ferent from ours generically only
gave results corresponding to $U(1)$ or $U(\infty)$ gauge groups,
see e.g.~\cite{Azuma:2004ie,Azuma:2005bj,Azuma:2004zq} and
references therein. The dynamical generation of a nontrivial index
on noncommutative spaces has also been observed in~\cite{Aoki:2004sd,Aoki:2006zi} for dif\/ferent models.

Our mechanism may also be very interesting in the context of the
recent observation~\cite{Abel:2005rh} that extra dimensions are very
desirable for the application of noncommutative f\/ield theory to
particle physics. Other related recent work discussing the
implications of the higher-dimensional point of view on symmetry
breaking and Higgs masses can be found in~\cite{Lim:2006bx,Dvali:2001qr,Antoniadis:2002ns,Scrucca:2003ra}.
These issues could now be discussed within a renormalizable
framework.

\section{Concluding remarks on the use of Fuzzy extra dimensions}

Non-commutative Geometry has been regarded as a promising framework
for obtaining f\/inite quantum f\/ield theories and  for regularizing
quantum f\/ield theories. In general quantization of f\/ield theories on
non-commutative spaces has turned out to be much more dif\/f\/icult and
with less attractive ultraviolet features than expected,
see however~\cite{Grosse:2004ik} and~\cite{Steinacker}.
Recall also that non-commutativity is not the only suggested tool for
constructing f\/inite f\/ield theories. Indeed four-dimensional f\/inite
gauge theories have been constructed in ordinary space-time and not
only those which are ${\cal N} = 4$ and ${\cal N} = 2$ supersymmetric,
and most probably phenomenologically uninteresting, but also chiral
${\cal N} = 1$ gauge theories \cite{Kapetanakis:vx} which already have
been successful in predicting the top quark mass and have rich
phenomenology that could be tested in future colliders
\cite{Kapetanakis:vx,zoup-kmz1,zoup-kmz2,zoup-kkmz1,zoup-acta,
Kobayashi:2001me,Babu:2002in}.
In the present work we have not addressed the f\/initeness of
non-commutative quantum f\/ield theories, rather we have used
non-commutativity to produce, via Fuzzy-CSDR, new particle models from
particle models on $M^4\times  (S/R)_F$.

A major dif\/ference between fuzzy and ordinary SCDR is that in the
fuzzy case one always embeds $S$ in the gauge group $G$ instead of
embedding just $R$ in $G$. This is due to the fact that the
dif\/ferential calculus on the fuzzy coset space is based on $\dim S$
derivations instead of the restricted $\dim S - \dim R$ used in the
ordinary one.  As a result the four-dimensional gauge group $H =
C_G(R)$ appearing in the ordinary CSDR after the geometrical
breaking and before the spontaneous symmetry breaking due to the
four-dimensional Higgs f\/ields does not appear in the Fuzzy-CSDR. In
Fuzzy-CSDR the spontaneous symmetry breaking mechanism takes already
place by solving the Fuzzy-CSDR constraints. The four-dimensional
potential has the typical ``mexican hat'' shape, but it appears
already spontaneously broken. Therefore in four dimensions appears
only the physical Higgs f\/ield that survives after a spontaneous
symmetry breaking. Correspondingly in the Yukawa sector of the
theory we have the results of the spontaneous symmetry breaking,
i.e.\ massive fermions and Yukawa interactions among fermions and the
physical Higgs f\/ield. Having massive fermions in the f\/inal theory is
a generic feature of CSDR when $S$ is embedded in $G$
\cite{Kapetanakis:hf}. We see that if one would like to describe the
spontaneous symmetry breaking of the SM in the present framework,
then one would be naturally led to large extra dimensions.

A fundamental dif\/ference between the ordinary CSDR and its fuzzy
version is the fact that a non-Abelian gauge group $G$ is not really
required in high dimensions. Indeed  the presence of a $U(1)$ in the
higher-dimensional theory is enough to obtain non-Abelian gauge
theories in four dimensions.


In a further development, we have presented a renormalizable four
dimensional $SU(N)$ gauge theory with a suitable multiplet of
scalars, which dynamically develops fuzzy extra dimensions that form
a fuzzy sphere. The model can then be interpreted as 6-dimensional
gauge theory, with gauge group and geometry depending on the
parameters in the original Lagrangian. We explicitly f\/ind the tower
of massive Kaluza--Klein modes, consistent with an interpretation as
compactif\/ied higher-dimensional gauge theory, and determine the
ef\/fective compactif\/ied gauge theory. This model has a unique vacuum,
with associated geometry and low-energy gauge group depending only
on the parameters of the potential.

There are many remarkable aspects of this model.  First, it provides
an extremely simple and geometrical mechanism of dynamically
generating extra dimensions, without relying on subtle dynamics such
as fermion condensation and particular Moose- or Quiver-type arrays
of gauge groups and couplings, such as in \cite{Arkani-Hamed:2001ca}
and following work. Rather, our model is based on a basic lesson
from noncommutative gauge theory, namely that noncommutative or
fuzzy spaces can be obtained as solutions of matrix models. The
mechanism is quite generic, and does not require f\/ine-tuning or
supersymmetry. This provides in particular a realization of the
basic ideas of compactif\/ication and dimensional reduction within the
framework of renormalizable quantum f\/ield theory. Moreover, we are
essentially considering a large $N$ gauge theory, which should allow
to apply the analytical techniques developed in this context.

In particular, it turns out that the generic low-energy gauge group
is given by $SU(n_1) \times SU(n_2) \times U(1)$ or $SU(n)$, while
 gauge groups which are
products of more than two simple components (apart from $U(1)$) do
not seem to occur in this model. The values of $n_1$ and $n_2$ are
determined dynamically. Moreover,  a magnetic f\/lux is induced in the
vacua with non-simple gauge group, which
 is very interesting in the
context of fermions, since internal f\/luxes naturally lead to chiral
massless fermions. This will be studied in detail elsewhere.

There is also an intriguing analogy between our toy model and string
theory, in the sense that as long as $a=0$, there are a large number
of possible vacua (given by all possible partitions)
corresponding to compactif\/ications, with no dynamical selection
mechanism to choose one from the other. Remarkably this analog of
the ``string vacuum problem'' is simply solved by adding a term to
the action.


\subsection*{Acknowledgments}
G.Z. would like to thank the organizers for the warm hospitality. This
work is supported by the EPEAEK programmes ``Pythagoras'' and co-funded
by the European Union (75\%) and the Hellenic state (25\%); also
supported in part by the mexican grant PAPIIT-UNAM IN115207.

\pdfbookmark[1]{References}{ref}
\LastPageEnding
\end{document}